\documentstyle[12pt,aaspp4]{article}
\def\singlespace{\def\baselinestretch{\@tightleading}\normalsize}
\begin{document}
\tighten

\title{Interstellar Matter and the Boundary Conditions of the Heliosphere\footnote{Talk at ACE Science Meeting, January 1997, Caltech}}

\author{Priscilla C. Frisch}
\affil{Department of Astronomy and Astrophysics, University of 
Chicago,\\
    Chicago, Illinois  60637}
\begin{center}
February 27, 1998
\end{center}

\begin{abstract}

The interstellar cloud surrounding the solar
system regulates the galactic environment of the Sun, and determines the boundary conditions of the heliosphere.
Both the Sun and interstellar clouds move through
space, so these boundary conditions change with time.
Data and theoretical models now support densities in the cloud 
surrounding the solar system of  n(H$^{\circ}$)=0.22$\pm$0.06 cm$^{-3}$,
and n(e$^{-}$)$\sim$0.1 cm$^{-3}$, with larger values allowed for
n(H$^{\circ}$) by radiative transfer considerations.
Ulysses and Extreme Ultraviolet Explorer satellite He$^{\circ}$ data yield a cloud temperature of {\mbox 6,400 K}.
Nearby interstellar gas appears to be structured and inhomogeneous.  
The interstellar gas in the Local Fluff cloud complex exhibits 
elemental abundance patterns in which refractory elements are
enhanced over the depleted abundances found in cold disk gas.
Within a few parsecs of the Sun, inconclusive evidence for factors of 2--5 variation
in Mg$^{+}$ and Fe$^{+}$ gas phase abundances is found, providing evidence
for variable grain destruction. 
In principle, photoionization calculations for the surrounding cloud 
can be compared with elemental abundances
found in the pickup ion and anomalous cosmic ray populations to
model cloud properties, including ionization, reference abundances, 
and radiation field.
Observations of the hydrogen pile-up at the nose of the heliosphere are consistent
with a barely subsonic motion of the heliosphere with respect to the surrounding
interstellar cloud.
Uncertainties on the velocity vector of the cloud that surrounds the solar
system indicate that it is uncertain as to whether the Sun and $\alpha$
Cen are or are not immersed in the same interstellar cloud.
\end{abstract}

\section{Introduction}

The physical conditions of the surrounding interstellar cloud
establish the boundary conditions of the solar system and heliosphere. 
The abundances and ionization
states of elements in the surrounding interstellar cloud determine the
properties of the parent population of the anomalous cosmic ray and
pickup ion components.  In addition, the
history of the interstellar environment of the heliosphere
appears to be partially recorded by radionucleotides such as $^{10}$Be and $^{14}$C in geologic ice core records
(\cite{sonett,fr97}).
Because the solar wind density decreases 
as $R^{-2}$ ($R$=distance to Sun), 
the solar wind and interstellar densities are equal at about 5 AU (the orbit of Jupiter), 
in the absence of substantial ``filtration'' \footnote{``Filtration'' refers to the deflection of interstellar H$^{\circ}$ around the heliopause due to the coupling between interstellar
protons and H$^{\circ}$ resulting from charge exchange}.
Approximately 98\% of the diffuse material in the heliosphere is
interstellar gas (\cite{gruntman}). Thus, the physical properties of the outer heliosphere are
dominated by interstellar matter (ISM).  Were the Sun to encounter
a high density interstellar cloud, it is anticipated that the physical properties of the inner
heliosphere would also be ISM-dominated.
Zank and Frisch (1998) have shown that if the space density 
of the interstellar cloud which surrounds the solar system were increased
to $\sim$10 cm$^{-3}$, the properties of the inner heliosphere at the
1 AU position of the Earth would be dramatically altered.

The accuracy with which the physical properties of the
surrounding cloud can be derived from observations of stars within a
few parsecs of the Sun (1 pc$\sim$200,000 AU) depends on the 
homogeneity and physical parameters of nearby ISM.  
Observations of nearby stars gives sightlines which probe the
ensemble of nearby clouds constituting the ``Local Fluff'' cloud complex.
Conclusions based on observations of nearby stars,  however, 
must be qualified by the absence of detailed data pertaining to the 
small scale structure of the local ISM (LISM).  More distant cold diffuse 
interstellar gas is highly structured, replete with dense ($\sim 10^{4}-10^{5}$ cm$^{-3}$), small (20--200 AU) 
inclusions occupying in some cases less than 1\% of the cloud volume (\cite{frail,falgarone,falpug,heiles}). 
Small scale structures are ubiquitous in interstellar gas, and individual
velocity components exhibiting column densities as low as N(H$^{\circ}$)$\sim$3$\times 10 ^{18}$ cm$^{-3}$ are found in cold clouds (\cite{frail,heiles}).
The presence of dense low column density 
wisps near the Sun is allowed by currently available data.

The Sun has a peculiar motion with respect to the ``Local Standard of Rest'' 
(LSR\footnote{The LSR is the velocity frame of reference in which the
vector motions of a group of nearby comparison stars are minimized.  
Stars in the LSR corotate around the galactic center with a velocity of
$\sim$250 km s$^{-1}$}); the Sun moves through the LSR
with a velocity V$\sim$16.5 km s$^{-1}$ towards the apex direction l=53$^{\circ}$, b=+25$^{\circ}$ (\cite{mihalas}).
Uncertainties on the relative solar-LSR motion appear to be less than 3 km s$^{-1}$
and $\pm$5$^{\circ}$.  This motion corresponds to $\sim$17 pc per million years.  Note that the solar path
is tilted by $\sim25^{\circ}$ with respect to the galactic plane.  The
Sun oscillates about the galactic plane, crossing the plane every 33 Myrs, 
reaching a maximum distance from the plane of $\sim$77 pc.  The last galactic plane
``crossing'' was about 21 Myrs ago (\cite{bash}).  
This amplitude of oscillation can be compared to
scale heights on the order of $\sim$50-80 pc for cold H$_{2}$ and CO, $\sim$100 pc
for cold H$^{\circ}$ and infrared cirrus, $\sim$250 pc for warm H$^{\circ}$, and
$\sim$1 kpc for warm H$^{+}$ (the ``Reynolds Layer'').

There are three time scales of interest in understanding the environmental
history of the solar galactic milieu -- $\sim 10^{6}$ years, $\sim 10^{5}$ years,
and $\sim 10^{4}$ years. 
Prior to entering into the Local Fluff complex of interstellar clouds, the Sun traveled
through a region of the galaxy between the Orion spiral arm and the spiral
arm spur known as the Local Arm.  On the order of a million years ago,
the Sun was displaced $\sim$17 pc in the anti-apex direction, which is
towards the present day location of the junction of the borders
of the constellations of Columba, Lepus and Canis Major.  
The motions of the Sun and surrounding interstellar cloud with respect
to interstellar matter within 500 pc, projected onto
the plane, are illustrated in Figure 1.  Note that the velocity
vectors of the Sun and interstellar cloud surrounding the solar system
are nearly perpendicular in the LSR, implying that the surrounding
cloud complex is sweeping past the Sun (see section \ref{velocity}).
When the morphology of the Local Fluff complex is
considered, it is apparent that sometime during the past $\sim$200,000 years the Sun appears to have emerged
from a region of space with virtually no interstellar matter (densities
n(H$^{\circ})<0.0005$ cm$^{-3}$, n(e$^{-})<0.02$ cm$^{-3}$) and entered the
Local Fluff complex of clouds (average densities n(H$^{\circ}$)$\sim$0.1 cm$^{-3}$)
outflowing from the Scorpius-Centaurus Association of star-forming regions.
One model for the morphology of the cloud surrounding the
solar system predicts that
sometime within the past 10,000 years, and possibly within the past 2,000
years, the Sun appears to have entered
the interstellar cloud in which it is currently situated (\cite{fr94}, Frisch 1997).
The cloud surrounding the solar system will be called here the ``Local Interstellar Cloud'' (LIC\footnote{This cloud
surrounding the solar system is also referred to as the ``surrounding interstellar cloud'',
or SIC, which unambiguously defines the cloud feeding interstellar matter
into the solar system.  For the sake of uniformity of notation, however,
the term LIC is used here.}).

\begin{figure}
\begin{center}
\plotone{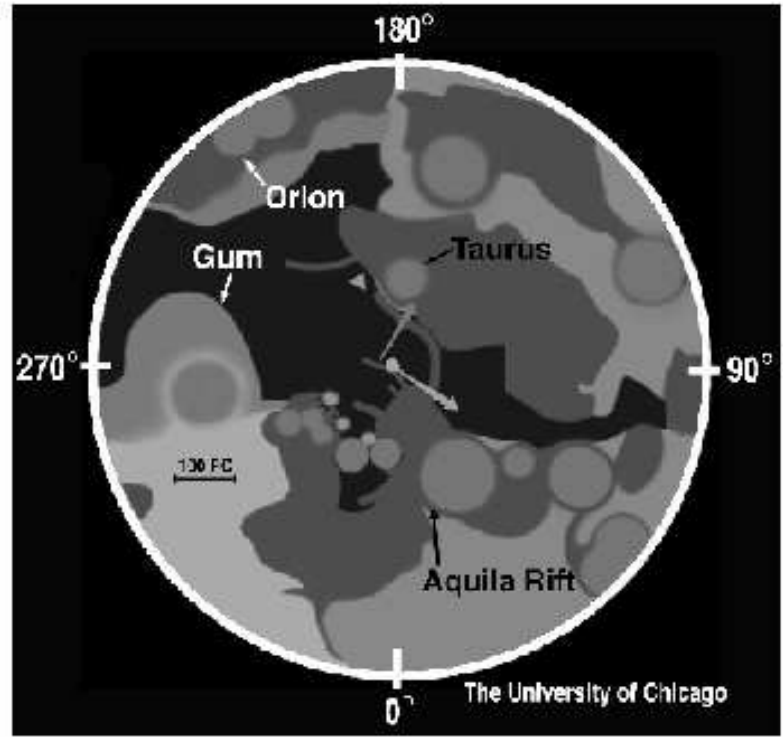}
\end{center}
\vspace{0.5in}
\caption[]{{\small
The distribution of interstellar molecular clouds (traced
by the CO 1-$>$0 115 GHz rotational transition) and diffuse gas (traced 
by E(B-V) color excess due to the reddening of starlight by interstellar dust) 
within 500 pc of the Sun are shown.  The round circles are molecular
clouds, and the shaded material is diffuse gas.  
The horizontal bar (lower left) illustrates a distance of 100 pc.  
Interstellar matter is shown projected onto the galactic plane, and the plot is labeled
with galactic longitudes.  The distribution of nearby interstellar matter
is associated with the local galactic feature known as ``Gould's Belt'',
which is tilted by about 15--20$^{\circ}$ with respect to the galactic plane.
ISM towards Orion is over 15$^{\circ}$ below the plane, while
the Scorpius-Centaurus material (longitudes 300$^{\circ}$--0$^{\circ}$) is about
15--20$^{\circ}$ above the plane.  Also illustrated are the space motions of the
Sun and local interstellar gas, which are nearly perpendicular in the 
LSR velocity frame.
The three asterisks are three subgroups of the
Scorpius-Centaurus Association.  The three-sided star is the Geminga
Pulsar.  The arc towards Orion represents the Orion's Cloak supernova
remnant shell.  The other arcs are illustrative of superbubble shells
from star formation in the Scorpius-Centaurus Association subgroups.  The
smallest (i. e. greatest curvature) shell feature represents the Loop I
supernova remnant.}}
\label{fig1}
\end{figure}

\section{Velocity, Magnetic Field\label{velocity}}

The velocity of the cloud surrounding the solar system provides the sole
criteria for selecting LIC absorption features in nearby stars, and 
therefore for deducing the structure of this cloud.

The velocity of the cloud feeding neutral gas into the solar system 
is found from backscattered solar He$^{\circ}$ 584 A radiation (e. g.
\cite{flynn}), direct measurements of inflowing
He$^{\circ}$ atoms by Ulysses (\cite{witte,witte2}, Witte, private
communication), and measurements of
pickup ions (\cite{moebius}).  
In the rest frame of the Sun, the downwind direction of the velocity vector found
from He$^{\circ}$ backscattered radiation is V=26.4$\pm$1.5 km s$^{-1}$, 
$\lambda$=76.0$^{\circ}$$\pm$0.4$^{\circ}$, $\beta$=--5.4$^{\circ}$$\pm$0.6$^{\circ}$ (ecliptic coordinates), with T=6,900$\pm$600 K (\cite{flynn}).
This corresponds to an heliocentric upwind vector in galactic coordinates 
of l=4.1$^{\circ}$$\pm$0.8$^{\circ}$ b=+14.8$^{\circ}$$\pm$0.7$^{\circ}$, V=--26.4$\pm$1.5 km s$^{-1}$.
The Ulysses neutral gas detector yields a downwind vector of V=25.3$\pm$0.5 km s$^{-1}$, $\lambda$=74.8$^{\circ}$$\pm$0.5$^{\circ}$, $\beta$=--5.7$^{\circ}$$\pm$0.2$^{\circ}$, T=6,100$\pm$300 K.  The Ulysses vector corresponds
to an heliocentric upwind direction in galactic
coordinates of l=3.8$^{\circ}$, b=+16.0$^{\circ}$.

These determinations of the cloud velocity from measurements of interstellar
gas within the solar system can be compared with
the values derived from observations of interstellar absorption lines towards the
nearest stars.
The velocity vector of the surrounding cloud has also been determined
by fitting the velocities of interstellar absorption line components in a
limited number of nearby stars.
The fits by Bertin et al. (1993a) give an upwind direction V=--25.7 km s$^{-1}$, l=5.9$^{\circ}$, b=16.7$^{\circ}$,
which is consistent with the Ulysses He$^{\circ}$ observations.
Uncertainties on this velocity are hard to ascertain due to the small number of stars
on which the vector derivation is based, but will be assumed
to be $\pm$1.5 km s$^{-1}$, or $\sim$1/2
of the typical spectral resolution of the data.

Removing solar motion from these three velocity vectors gives an 
upwind direction for the LIC cloud in the LSR of V=--18.7$\pm0.6$ km s$^{-1}$ 
from the direction l=327.3$^{\circ}$$\pm$1.4$^{\circ}$, b=0.3$^{\circ}$$\pm$1.0$^{\circ}$ (\cite{fr95}).
The Ulysses and EUVE data He$^{\circ}$ data indicate a cloud 
temperature of T=6,400 K.  As a curiosity, this LSR direction is $\sim$10$^{\circ}$ from the 1.3 pc distant star $\alpha$ Cen.  In essence, the Local
Fluff gas is sweeping down on the solar system from the vicinity
of $\alpha$ Cen.

The small difference in the magnitudes and directions of the Ulysses versus backscatter vectors
introduces a significant uncertainty concerning the structure of the Local
Fluff and the location of the Sun with respect to
the boundary of the surrounding interstellar cloud.
The uncertainties on the velocity vectors as defined by the Flynn et al.
versus Ulysses versus Bertin et al. (``AG'') vectors do not allow
us to make a definitive statement as to whether or not the Sun
and the nearby star $\alpha$ Cen are immersed in the same
interstellar cloud.
The Flynn et al. vector projected in the direction of the star $\alpha$ Cen (d=1.3 pc, l=315.8$^{\circ}$, b=--0.7$^{\circ}$)
predicts a heliocentric velocity of --16.9$\pm$1.0 km s$^{-1}$ for the LIC cloud.
The Ulysses flow vector predicts a velocity of --16.2$\pm$1.0 km s$^{-1}$.
The Bertin et al. AG flow vector predicts a
velocity of --15.7$\pm$1.0 km s$^{-1}$.  The small differences between
these three predictions ($\sim$1.2 km s$^{-1}$) may seem academic,  
but when compared to observational data these differences have consequences
regarding the location of the Sun with respect to the edge of the
interstellar cloud in which it is situated.
The velocities of the
Mg$^{+}$ and Fe$^{+}$ absorption lines observed in this direction 
are --17.5 to --18.2 km s$^{-1}$ with assumed uncertainties on the individual
velocities of $\pm$1.0 km s$^{-1}$ (corresponding to 1/3 of the nominal
spectral resolution with which the observations were acquired, \cite{lal95,linwood97})
\footnote{These data were acquired with the Hubble Space Telescope GHRS instrument at a nominal spectral resolution of 3.3 km s$^{-1}$.}.
As noted by Lallement et al. (\cite{lal95}), if the AG velocity
vector is correct, less than 2.5 x 10$^{16}$ cm$^{-2}$ column density
of H
is found between the Sun and the location of the cloud edge in this direction.
This corresponds to a distance to the
cloud edge of less than 0.03 pc (6000 AU) for n(H)=0.3 cm$^{-2}$ (see section \ref{neutral}).  In contrast, if the Flynn et al. vector is
correct, the uncertainties allow the Fe$^{+}$ and Mg$^{+}$ absorption
features to be formed in the surrounding cloud,
so that log(N(Fe$^{+}$))=12.36 cm$^{-2}$ towards $\alpha$ Cen (\cite{lal95}).
Using log(N(Fe$^{+}$)/N(H$^{\circ}$))$\sim$--5.71 in the surrounding cloud
(Table 1) then indicates that the cloud edge is at or beyond $\alpha$ Cen.
Since the star $\alpha$ Cen is located in the true upwind direction
(i. e. in the LSR), if the AG vector is correct the Sun would thus exit the cloud now
surrounding the solar system within the next 2,000 years, while if the
Flynn et al. vector is correct the Sun will not exit the current cloud 
for over 70,000 years.
These data allow, but do not require, the Local Fluff gas to be a structured or turbulent medium
on the $\leq$5,000 AU scale. 
Evidently reducing these uncertainties in the velocity of the cloud
around the solar system will enable us to understand
interactions between the Sun and interstellar gas over the next 100,000 years.

The strength of the interstellar magnetic field in the cloud 
surrounding the solar system has not been directly measured.  
A strength of $\sim$1.6 $\mu$G has been found to be consistent with 
observations of polarization and pulsar dispersion measures
(\cite{fr91}), while estimates of the heliosphere confinement
pressure give an upper limit B$<$3--4 $\mu$G (\cite{gloeckler}).
Ultimately, three-dimensional MHD heliosphere models, in comparison
with {\it Voyager} data, may provide the best estimate for the
magnetic field strength in the cloud surrounding the solar system.
The presence of a local magnetic field is confirmed by observations of the polarization
of light from nearby stars (d$<$35 pc) in the galactic center 
hemisphere (\cite{tinb,fr91}).  {\it If} the cloud surface is perpendicular
to the direction of gas flow, then the direction of the magnetic
field is found to be approximately parallel to the LIC cloud surface (\cite{fr96}).
The pile-up of H$^{\circ}$ at the heliosphere nose is consistent with an interstellar magnetic field strength of 3 $\mu$G (see section \ref{wall}).

\section{Neutral Density\label{neutral}}

The quantities of interest in constraining the heliosphere are the absolute neutral and ion
space densities (cm$^{-3}$), which can be derived from column densities
(cm$^{-2}$) only if cloud length is known.  
The {\it average } density of nearby gas can be 
inferred from observations of H$^{\circ}$ and other elements seen in absorption.
For example, combining H$^{\circ}$ column densities for the six clouds detected
towards the four nearby stars $\alpha$ Cen (1.3 pc),
$\alpha$ CMa (2.7 pc), $\alpha$ CMi (3.5 pc) and $\alpha$ Aql (5 pc), gives an average density $<$n(H$^{\circ}$)$>$=0.08--0.11 cm$^{-3}$ for interstellar
gas within 3--5 pc of the Sun. 
This value can be compared with $<$n(H$^{\circ}$)$>$$\sim$0.2 cm$^{-3}$ for the surrounding cloud, suggesting a filling factor of $\sim$50\%\footnote{The 
``filling factor'' of a cloud complex is the fraction of space occupied by the gas in the complex.}.
  Data for $\alpha$ Cen and $\alpha$ CMa are given
in Table 1.  The hydrogen column density for $\alpha$ Aql has been inferred from
Ca$^{+}$ absorption lines using 
the conversion factor N(Ca$^{+}$)/N(H)$\sim$10$^{-8}$,
based on N(Ca$^{+}$)$\sim$10$^{10}$ cm$^{-2}$ and N(H$^{\circ}$+H$^{+}$)=10$^{18}$ cm$^{-2}$
towards $\eta$ UMa (which samples a low column density sightline through the Local Fluff complex, \cite{fw97}).  
With this conversion factor, 
N(Ca$^{+}$)$\sim$1.7 $\times$ 10$^{10}$ observed towards $\alpha$ Aql,
gives N(H)=1.7 $\times$ 10$^{18}$ cm$^{-2}$.
(The depletion variations of Ca$^{+}$, discussed in section 5, will affect this
conclusion.)
The average density $<$n(H$^{\circ}$)$>$$\sim$0.10 cm$^{-3}$ 
can be used as a guide when inferring the overall morphology of interstellar clouds in the
Local Fluff complex.

The absolute space density of the LIC can be inferred from observations of
pickup ions combined with the H$^{\circ}$/He$^{\circ}$ ratio in the 
surrounding cloud (e. g. \cite{gloeckler}, \cite{val96,fs96}).
An average value N(H$^{\circ}$)/N(He$^{\circ}$)$\sim$14 is found from observations of white dwarf stars (\cite{dup95,fr95,val96}, also see section \ref{abundances}).
EUVE observations of the two lowest column density white dwarf stars
GD 71 (l=192$^{\circ}$, b=--5$^{\circ}$, N(H$^{\circ}$)=8.7$\pm$0.7 $\times$ 10$^{17}$ cm$^{-2}$) and HZ 43 (l=54$^{\circ}$, b=84$^{\circ}$, N(H$^{\circ}$)=6.3$\pm$1.6 $\times$ 10$^{17}$ cm$^{-2}$)
give N(H$^{\circ}$)/N(He$^{\circ}$) ratios of 12.1 and 15.8 respectively
over the sightlines to these stars
(\cite{val96}), a difference which may reflect local variations in ionization
levels.
In principle hydrogen ionization rises towards cloud exterior more rapidly
than helium ionization because of the low column densities
(\cite{sla89,chbr,val96,sf97a,sf97b}), so the actual ratio N(H$^{\circ}$)/N(He$^{\circ}$)
at the solar location will be larger than the sightline averaged values
measured for white dwarf stars.  
Observations by Ulysses of interstellar helium within the
heliosphere yield n(He$^{\circ}$)$\sim$0.016$\pm$0.002 cm$^{-3}$ (\cite{witte,witte2}, Witte
private communication).  
Combining the Ulysses and EUVE data gives a lower limit on the neutral density in the interstellar cloud outside of the solar system 
of n(H$^{\circ}$)=0.22$\pm$0.06 cm$^{-3}$, with larger values allowed
by the above radiative transfer considerations.
Both hydrogen and helium are ionized in the surrounding cloud, however, so 
the total space density (neutral plus ions) will be larger (see section \ref{ion}).

Observations of L$\alpha$ backscattered radiation from interstellar
hydrogen in the inner heliosphere yield a value n(H$^{\circ}$)=0.15$\pm$0.05 cm$^{-3}$
for the interstellar H$^{\circ}$ density inside of the heliopause (\cite{quem}).
Combined with the LIC value of n(H$^{\circ}$) above, this gives
$\sim$30\% filtration of H$^{\circ}$ at the heliopause. 

\section{Structure\label{structure}}

The largest data set on the Local
Fluff complex is derived from observations of the optical Ca$^{+}$
H and K lines.  Although globally calcium has highly variable depletions
onto interstellar dust grains in the ISM, Ca$^{+}$ observations
constitute the most complete set of available data and therefore
provide a first look at the structure and morphology of nearby gas.
The rest of this section rests on the assumption that calcium depletions
and ionization are relatively constant across the clouds in the 
Local Fluff cloud complex.  This premise may not be true, but it 
offers an insight into the distribution of ISM within 30 pc of the Sun.
Because of the variation of both calcium depletion and ionization in interstellar gas in general, this distance estimate is useful mainly to provide a qualitative
picture of the distribution of Local Fluff gas.

About 80 individual Ca$^{+}$ velocity components have been observed 
that can be attributed to the Local Fluff complex (\cite{fw97}).
The overall morphology of the Local Fluff complex is shown in Figure 2,
where Ca$^{+}$ column densities (N(Ca$^{+}$)) are plotted
versus galactic longitude and latitude 
for stars sampling the Local Fluff complex.
(Data are from 
\cite{ber93b,c94,cd95,ccw97,fr97,lal86,lb92,lal94,lal95,val93,wel96,fw97}.)
The well-known asymmetry in the distribution of Local Fluff
complex between the galactic center and anti-center
hemispheres is apparent.
In the absence of data on non-refractory elements in a given sightline, 
the distance to the ``edge'' of the Local Fluff cloud complex 
can be guesstimated using the somewhat uncertain relation for the Local Fluff  
d(pc)$\sim$3 $\times$ 10$^{-10}$ N(Ca$^{+}$), for $<$n$>$$\sim$0.1 cm$^{-3}$.
The sightline to the star $\tau^{3}$ Eri (l=214$^{\circ}$, b=--60$^{\circ}$,
N(Ca$^{+}$)=5 $\times$ 10$^{10}$ cm$^{-2}$) shows that the
Local Fluff complex extends into the southern galactic hemisphere.
The low column densities (N(Ca$^{+}$)$<$10$^{10}$ cm$^{-2}$, \cite{val93,fw97}) 
towards the three high latitude stars 
$\sigma$ Leo (l=253$^{\circ}$, b=60$^{\circ}$), $\eta$ UMa (l=101$^{\circ}$, b=65$^{\circ}$), and $\alpha ^{2}$ CVn (l=117$^{\circ}$, b=79$^{\circ}$)
show the Sun is ``above'' most of the mass of the Local Fluff
complex (in the sense of ``above'' the galactic plane).  The cloud in front of $\tau ^{3}$ Eri may be an extension of the
interstellar cloud complex
seen towards $\alpha$ Gru, $\delta$ Vel, $\epsilon$ Gru, $\alpha$ Hyi, 
and $\iota$ Cen and other nearby stars (with data from \cite{cd95,lal86},
\cite{ccw97,fw97}).
The location of the upwind cloud in the LSR (which is the galactic center
hemisphere of the sky) coincides with the interstellar dust patch 
which polarizes the light of nearby stars (\cite{tinb}, \cite{fr91}). 
This cloud has been identified as part of the leading edge of the 4 Myr 
old superbubble shell expanding from the last epoch of star-formation 
in the Scorpius-Centaurus association (\cite{fr95}).
\begin{figure}[t!]
\vspace*{1.0in}
\plotone{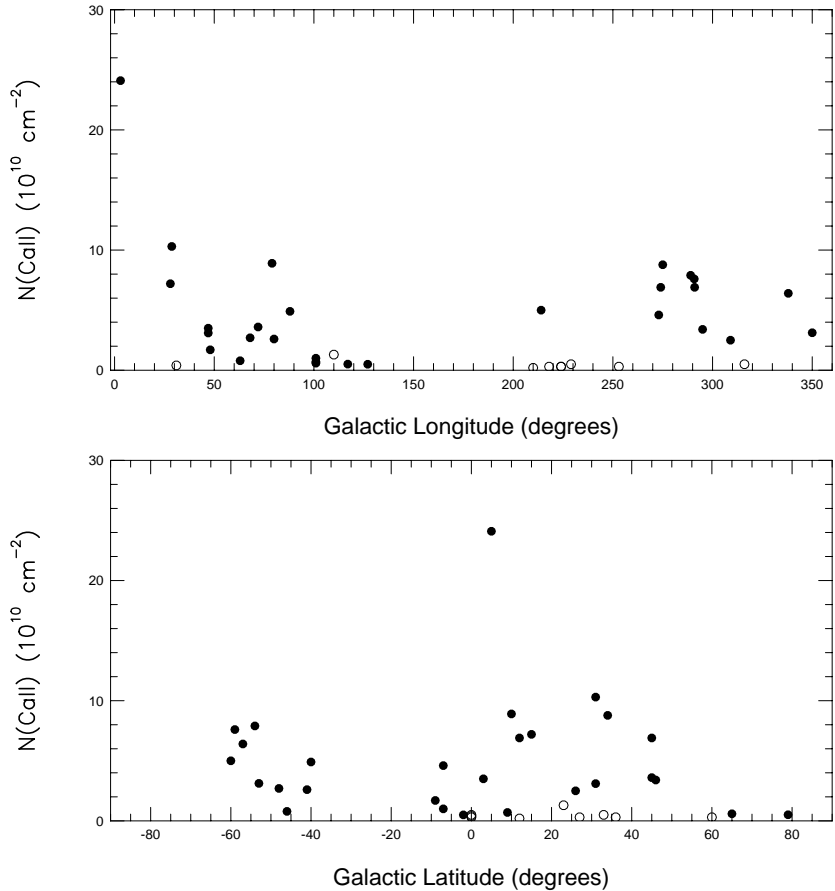}
\vspace{0.8in}
\caption[]{\small
Overall morphology of Local Fluff complex of interstellar clouds based on
Ca$^{+}$ observations of nearby stars. Total Ca$^{+}$ column densities 
for stars sampling the Local Fluff complex are
plotted versus galactic longitude and latitude. 
The open circles are upper limits.
With the exception of $\tau$$^{3}$ Eri, at l=214$^{\circ}$, b=--60$^{\circ}$, d=17 pc,
the section of the galaxy between l=130$^{\circ}$ and 270$^{\circ}$ is 
relatively devoid of nearby interstellar gas.
(Data are from \cite{ber93b,c94,cd95,ccw97,fr97,lal86,lb92,lal94,lal95,val93,wel96}.)
}
\label{NCa_lb.eps}
\end{figure}

The Local Fluff complex is structured and inhomogeneous.
At 3 km s$^{-1}$ resolution, about one interstellar absorption line velocity component
is seen per 1-2 parsecs of sightline in the direction of the stars
$\alpha$ CMa at 2.7 pc and $\alpha$ Aql at 5 pc (\cite{lal94}, \cite{lal86}),
but the actual filling factors of the gas is not known since the densities
of LISM clouds other than the LIC are unknown.
Globally, cold H$^{\circ}$ clouds and cold molecular (H$_{2}$, CO) clouds fill $\sim$10$^{-3}$--10$^{-4}$ of cloud volume, 
evidently revealing structure consistent with a Kolmogorov spectrum, including
features on $\leq$100 AU scale sizes.  
Ionization also appears to be inhomogeneous, based on both the variations
in the white dwarf N(H$^{\circ}$)/N(He$^{\circ}$) ratios and factors of 2--4 
variation in LIC electron densities (section \ref{ion} and Table 1).   
Other possible evidence for local structure on subparsec scales
are the absence of gas at the LIC velocity in the Fe$^{+}$
and Mg$^{+}$ lines towards $\alpha$ Cen (1.3 pc) (section 2) and the possible
small scale structure ($\leq$0.1 pc)
revealed by the finite time width of the $^{10}$Be spikes in the
antarctic ice core record (\cite{fr97}).  The upper limits on Ca$^{+}$ column
density towards $\lambda$ Aql (l=30$^{\circ}$, b=--6$^{\circ}$, \cite{val93}) indicate
either the LIC cloud surface is closer than
1.2 pc or that the calcium is doubly ionized in this direction.

\section{Ionization\label{ion}}

The relative ionizations of elements in interstellar gas at the solar location 
are a sensitive function of the relative number of EUV (200--900 A)
versus FUV (900--1500 A) photons, which in turn depends on the mix of 
different sources of ionizing radiation and radiative transfer in the
Local Fluff (e. g. \cite{sla89,rey,chbr,valwel}).
Locally, hydrogen ionization is dominated by 
the ultraviolet flux from the star $\epsilon$ CMa, a luminous B2 II star
located in a direction with very low column densities.
The radiation field which ionizes He, Ne, and other
species with ionization potentials greater than \mbox{13.6 eV} is dominated by white dwarf
stars, the diffuse soft X-ray flux, and radiation from a predicted conductive
interface between the Local Fluff complex and the nominal hot 
($\sim$10$^{6}$) plasma filling the adjacent Local Bubble interior (Vallerga 1998, \cite{sf97a,sla89}).  

The question of which methods to use in calculating interstellar electron
densities needs to be put in the context of global ISM studies.
Different estimates of interstellar ionization equilibrium 
towards denser clouds do not produce consistent results.
Welty et al. (1998) have observed an n$\sim$10 cm$^{-3}$ cloud towards the star 23 Ori,
and found that thirteen different ratios of neutral-to-first ionization
states yield electron density values that span a range of a factor of 25.
Welty attributes this range to a combination of 
unidentified physical processes (such as charge exchange) and uncertain
rate constants.
In light of these generic ISM studies, for the LIC a variety of approaches are needed to estimate the
electron density.

The electron density in the LIC cloud has been found with several methods,
some of which provide cloud-averaged values versus values at the 
solar location.  Based on the results outlined below, it appears
that n(e$^{-}$)$\sim$0.1 cm$^{-3}$ in the LIC, but
more work is needed.  These methods include:

\begin{itemize}

\item Observations of absorption lines of trace ions in the LIC cloud as seen
towards nearby stars.  These data give electron density values
that are averaged over the cloud segment being sampled.

\item  Theoretical predictions of the LIC ionization at the solar
location based on a radiation field which is the sum of
measurements of ultraviolet and extreme ultraviolet fluxes of point sources
such as white dwarf and other hot stars near the Sun.  These calculations
include radiative transfer effects.

\item Theoretical predictions of the LIC ionization, including
radiative transfer effects, at the solar
location based on a radiation field which is the sum of
both a diffuse radiation field (from scattered starlight, plasma in
the local hot bubble, and emission from the predicted conductive
interface on the LIC cloud) and
measurements of ultraviolet and extreme ultraviolet fluxes 
from point sources such as white dwarf and hot stars near the Sun.

\item  Measurements of H$^{\circ}$, He$^{\circ}$, and He$^{+}$
in nearby white dwarf stars, which when compared to the nominal
reference abundance ratio N(H$^{\circ}$)/N(He$^{\circ}$=10/1 gives 
an estimate of the LISM ionization.  This approach gives electron 
densities averaged over the cloud segment being sampled.

\item  Theoretical models of the hydrogen wall of the heliosphere, 
which has properties dominated by charge exchange between interstellar protons
and hydrogen atoms piled up at the heliopause.  Proton and electron densities are assumed to
be equal.

\end{itemize}

Absorption Line Data: The LIC electron density has been inferred from the ratio Mg$^{+}$/Mg$^{\circ}$
and C$^{+}$ fine-structure lines (\cite{fw90,lal94,lalfer96,gry95,fr94},
\cite{lalissi,wlin97}).
Electron densities found towards nearby stars are 0.09 (+0.23,--0.07) cm$^{-3}$ towards 
$\epsilon$ CMa (LIC cloud only), 0.11 (+0.12,--0.06) towards $\alpha$ Aur,
$\sim$0.24 towards $\eta$ UMa,  0.31$\pm$0.20 
towards $\delta$ Cas, and 0.2--0.4  cm$^{-3}$ towards $\alpha$ CMa (see Table 1).
The large value seen towards $\alpha$ CMa may indicate a cloud interface
between the LIC cloud (T$\sim$10$^{4}$ K) and adjacent plasma (T$\sim$10$^{6}$ K, \cite{fr94,ber95}).
The convergence between the theoretical models (Table 2) and the
n(e$^{-}$) values from $\epsilon$ CMa
Mg$^{+}$/Mg$^{\circ}$ and $\alpha$ Aur C$^{+}$ fine-structure data
suggest that the LIC value is n(e$^{-}$)$\approx$0.1 cm$^{-3}$.
In principle, electron densities can also be inferred from
optical observations of Na$^{\circ}$ and Ca$^{+}$,
but as these species are subordinate ionization
states of elements with variable depletions in the ISM, and trace
ISM at different temperatures, the resulting electron densities are
highly uncertain.  The range n(e$^{-}$)=0.15$\pm$0.11 cm$^{-3}$ is found from 
Na$^{\circ}$ data (\cite{lalfer96,lalissi}).

Stellar radiation field:  Predicted ionization levels of the LISM depend on whether the
radiation field estimates are based on point sources, such as
white dwarf and hot stars, or also emission from diffuse hot gas. 
Including stellar fluxes alone, e. g. $\epsilon$ CMa and white dwarf stars,
hydrogen is predicted to be $\sim$13\% ionized and He $\sim$4.3\% ionized at
the solar location (\cite{val96,val98}), yielding n(e$^{-}$)$\sim$0.03 cm$^{-3}$

Diffuse and stellar radiation field:  Diffuse EUV emission is
expected from the
Local Bubble 10$^{6}$ K plasma and the conductive interface between the LIC and plasma (\cite{sla89}).  
As part of an ongoing project, we (Slavin and Frisch 1998a, 1998b, Frisch and Slavin 1996) are modeling the ionization balance in the LIC using
a radiative transfer ionization which includes both
diffuse and point source EUV and FUV emission, yielding a self-consistent
solution for cloud heating and cooling.  The predictions of this model
are sensitive to assumptions about the input radiation field and cloud
abundances.  
These results have been calculated using parts per million (PPM)
values for He, C, N, O and Ne of 1 $\times$ 10$^{5}$, 191, 85, 646, and 123,
respectively, based on LIC abundances towards $\epsilon$ CMa (\cite{grydup95}).
In this particular model hydrogen and helium ionizations are predicted to be $\sim$23\% 
and $\sim$36\%, respectively, giving n(e$^{-}$)$\sim$0.08 cm$^{-3}$ 
if n(H)=0.3 cm$^{-2}$.  Table 2 shows that these results for He, O, N are
consistent with PUI data providing O and N are filtered at the heliopause
by $\sim$30\%.  (Alternatively, they could have lower intrinsic abundances.)
However, Ne predictions agree with the PUI data only if Ne$^{\circ}$ 
increases by a factor of $\sim$2.5.  Since Ne is only $\sim$20\% neutral
at the solar location in this particular model, small changes in the 
extreme ultraviolet radiation field yield larger changes in the fraction
of neutral Ne.  Since He has about the same ionization potential, however,
this change would create problems in understanding the He data.
The model used for the Table 2 predictions yields predictions of
S$^{\circ}$/O$^{\circ}$=2.8 $\times$ 10$^{-6}$ and Mg$^{\circ}$/O$^{\circ}$=2.2 $\times$ 10$^{-5}$.
These predictions are highly sensitive to the diffuse radiation field in the
200--1200 A region which has not been directly measured.
These results represent one possible model, but better understandings of the
radiation field, ISM reference abundances, and cloud geometry are needed
before the problem can be considered solved.

White dwarf stars:  The two white dwarf stars with the lowest
column densities are GD 71 and HZ 43.  
EUVE observations of He$^{\circ}$ and He$^{+}$ toward the white dwarf star 
HZ 43 (a high latitude white dwarf at 60 pc which samples the Local
Fluff complex) yield an upper limit on the electron density
of n(e$^{-}$)$<$0.14 cm$^{-3}$ for n(H$^{\circ}$+H$^{+}$)=0.15--0.34 cm$^{-3}$,
after comparing N(H$^{\circ}$), N(He$^{\circ}$) and N(He$^{+}$) with
the expected cosmic ratio N(H)/N(He)=10/1 and heliospheric He data (\cite{val96}).

Hydrogen wall:  The observed velocity distribution of interstellar H$^{\circ}$ decelerated by charge exchange with
interstellar protons at the heliopause can be reproduced theoretically in the direction
of $\alpha$ Cen for the case where the interstellar gas confining the
heliosphere has plasma density
n(e$^{-}$)=0.1 cm$^{-3}$ (see section \ref{wall}).

\section{Abundances\label{abundances}}

In principle {\it in situ} measurements of pickup ions and 
anomalous cosmic rays can help resolve an active debate as to
whether solar or B-star abundances (which are $\sim$ 70\% solar)
apply as the correct reference abundance for the interstellar medium.\footnote{The abundances of
C, N, O and other elements in the Sun are $\sim$50\%  higher than abundances
 in B-stars,
and it is an open question as to whether B-star or solar abundances apply
to the ISM (\cite{snow,savsem}).}
Since He is primordial, the correct reference ratio H/He ratio should
$\sim$10/1, regardless of whether solar or B-star abundances apply in the ISM.
Elements that will be sensitive as to whether the correct ISM reference abundances
are solar versus B-star are Ne, Ar, N and S, none of which are
thought to be depleted onto interstellar dust grains.
Therefore, when LIC ionization and the filtration of
elements at the heliopause are properly understood, measurements
of N/He, O/He, Ne/He and Ar/He (for example) will constrain the
reference abundance of the LIC cloud.

The gas phase abundances of the neutral atomic constituents in the LIC 
are a function of both the ionization and depletion patterns of the gas.
Abundant elements in the ISM such as C, S, Mg, Fe, and Si, have
first ionization potentials (FIP) less than the 13.6 eV ionization potential
of H$^{\circ}$, so they will be predominantly ionized in the LIC.
Thus, low-FIP elements will be under-represented in the pickup ion
population when compared to cosmic abundances.
The relative abundances of common low-FIP elements,
based on observations typical of warm and cold gas in interstellar clouds, are shown
in Table 3.  Absorption lines from C$^{\circ}$, Si$^{\circ}$, Fe$^{\circ}$,
S$^{\circ}$ are weak in warm low column density ISM.

\begin{figure}[th]
\vspace*{8.5in}
\includegraphics{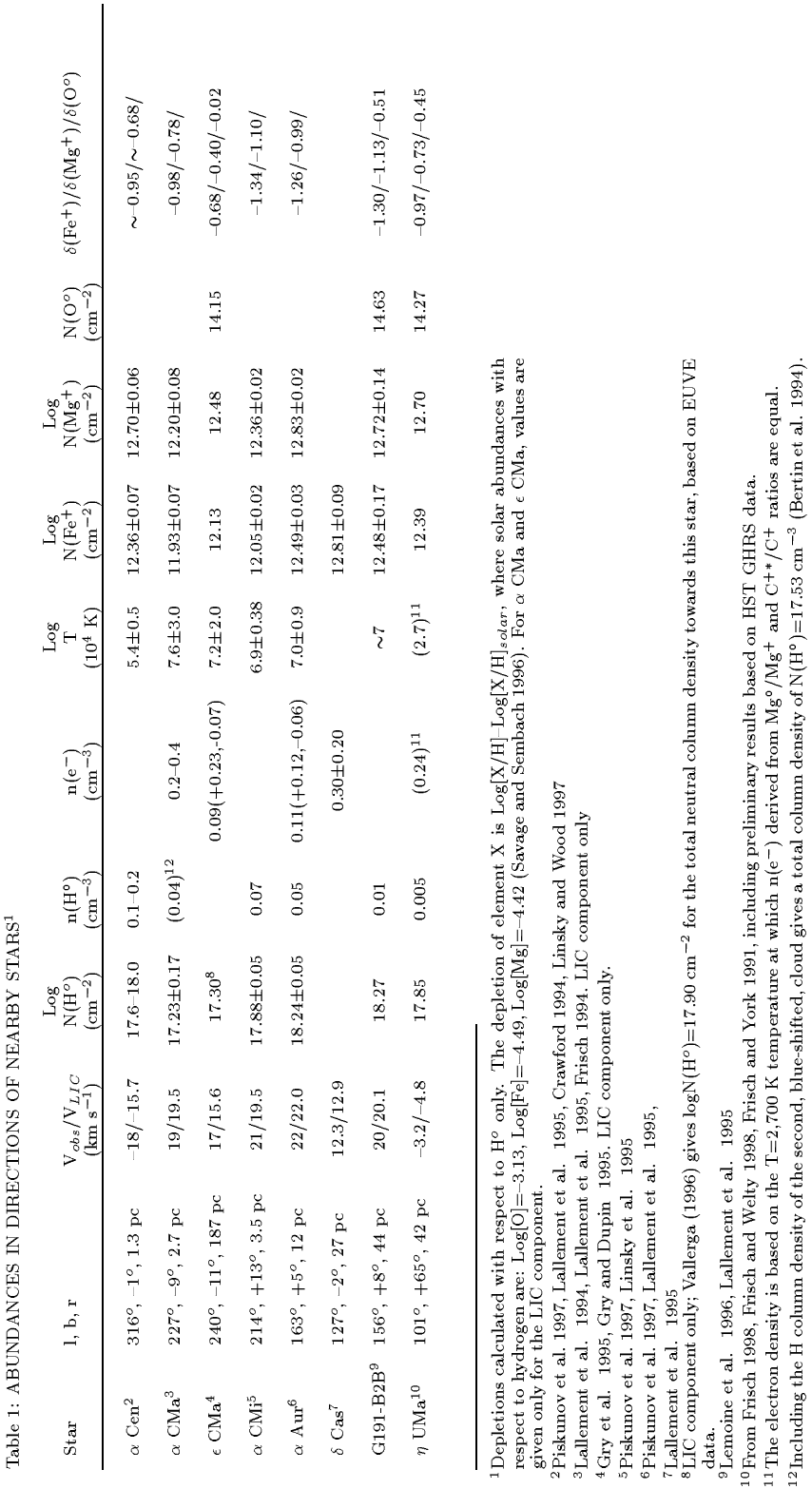}
\end{figure}
\clearpage

\clearpage
\begin{figure}[t!h]
\includegraphics{table2.ps}
\end{figure}
\clearpage

\begin{table}[t]
\large
\begin{center}
\normalsize
Table 3:  Gas-Phase Neutral Elements in Sample of Interstellar Clouds\\
\vspace*{0.5cm}
\normalsize
\begin{tabular}{c|c|c|c} \hline
Ratio & Log(Ratio)/H$^{\circ}$) & Source&Reference\\
\hline
O$^{\circ}$/H$^{\circ}$ & --3.19 & $\epsilon$ CMa LIC& Gry and Dupin (1995)\\
Mg$^{\circ}$/H$^{\circ}$ & --4.82& $\epsilon$ CMa LIC& Gry and Dupin (1995)\\
D$^{\circ}$/H$^{\circ}$ & --5.12 & $\lambda$ Sco warm cloud & York (1983)\\
N$^{\circ}$/H$^{\circ}$ & --4.14 & $\lambda$ Sco warm cloud & York (1983)\\
O$^{\circ}$/H$^{\circ}$ & --3.37 & $\lambda$ Sco warm cloud & York (1983)\\
Ar$^{\circ}$/H$^{\circ}$ & --5.81 & $\lambda$ Sco warm cloud & York (1983)\\
Na$^{\circ}$/H$^{\circ}$ & --8.06 & $\zeta$ Oph warm cloud & Morton (1975)\\
Mg$^{\circ}$/H$^{\circ}$ & --8.01 & $\zeta$ Oph warm cloud & Savage et al. (1992)\\
C$^{\circ}$/(H$^{\circ}$+H$_{2}$) & --5.77 & $\zeta$ Oph cold cloud & Morton (1975)\\
Si$^{\circ}$/(H$^{\circ}$+H$_{2}$) & $\leq$--8.54 & $\zeta$ Oph cold cloud & Morton (1975)\\
Fe$^{\circ}$/(H$^{\circ}$+H$_{2}$) & --9.6 & $\zeta$ Oph cold cloud & Morton (1975)\\
S$^{\circ}$/(H$^{\circ}$+H$_{2}$) & --7.19 & $\zeta$ Oph cold cloud & Federman and Cardelli (1995)\\
C$^{+}$/H$^{\circ}$ &--3.85& global ISM & Cardelli et al. 1998 \\
N$^{\circ}$/H$^{\circ}$ &--4.12& global ISM & Meyer et al. 1998a \\
O$^{\circ}$/H$^{\circ}$ &--3.50 & global ISM & Meyer et al. 1998b \\
\hline \\
\end{tabular}
\end{center}
\end{table}
\normalsize
\noindent

Most elements are not present in solar abundances in interstellar gas, 
a phenomena explained by invoking the depletion of the missing elements onto dust
grains which are mixed into the interstellar gas.   
Recently arguments have been made that the correct reference abundance for 
the interstellar medium should be B-star abundances, at about 70\% solar 
(e. g. \cite{snow}), reducing the amount of ISM tied up in grain mass.

Typical gas-to-dust mass ratios
of 100:1 are found in the global ISM.  Depletion patterns depend on cloud type, 
and nearby interstellar gas shows enhanced abundances with respect to
the abundance patterns seen in distant cold clouds (Frisch 1981).  
Generally elements with high condensation temperatures (e. g. Si, Mg, Mn, Fe, 
Cr, Ni, Ca, Co, Ti) are the most depleted in cold clouds.
In warm interstellar clouds these refractory elements show less depletion,
i. e. have higher gas phase abundances (e. g. see Savage and Sembach 1996).
Warm clouds in the galactic halo have similar properties as warm disk clouds
(e. g. \cite{fitzpatrick}).
The abundances of Fe and Mg in Table 1 indicate LISM gas-phase
abundance variations of factors of 2--5, suggesting variable grain destrution.
In the cold cloud towards the star $\zeta$ Oph, 99.4\% of
Fe and 97\% of Mg are depleted onto dust grains.
In contrast, $\sim$90\% of Fe and $\sim$85\% of Mg are depleted onto dust 
grains in the LIC (Table 1).
These enhanced abundances seen in local gas were the basis for my original
conclusion that the interstellar gas surrounding the solar system
had been processed through a supernova shock front
which had partially destroyed embedded dust grains (\cite{fr81}).

Figure 3 shows gas phase abundances of the most common elements,
where in this case the solar abundances have been used as the reference abundance
for interstellar gas.
The gas phase abundances in both warm and cloud clouds are shown.
Elements with FIPs less than 13.6 eV will be mainly ionized, whereas O, N, Ar, Ne, He will be
primarily neutral when hydrogen is neutral.  Also shown are uncertainties in 
Mg, Fe, Si and O abundances in the LIC.
Predicted ratios Mg$^{\circ}$/O$^{\circ}$=2.2 $\times$ 10$^{-5}$ and S$^{\circ}$/O$^{\circ}$=2.8 $\times$ 10$^{-6}$ 
are deduced from the radiative transfer model
with LIC abundances discussed in section \ref{ion} and 
Table 2.  Thus, Mg$^{\circ}$ and S$^{\circ}$ are candidates for positive 
detections in the pickup ion population.

\begin{figure}[t]
\vspace{3in}
\includegraphics{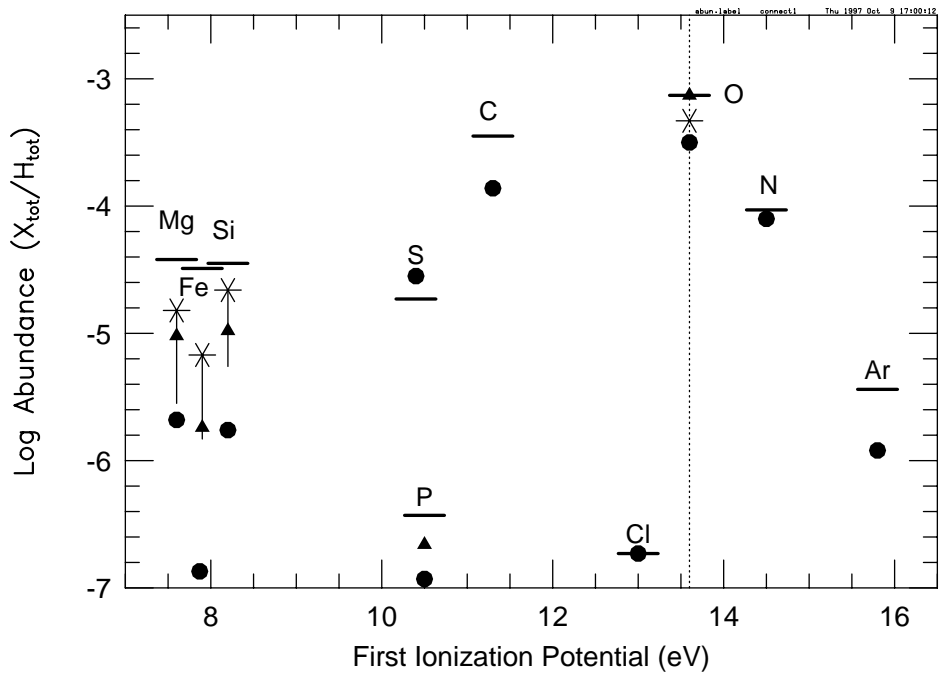}
\caption[]{\small
The abundances of the most common elements present in interstellar gas are
shown plotted against FIP.
X$_{tot}$ is the total abundance of element X, while H$_{tot}$
is the total abundance of H.
The solar abundances (short line), and typical abundances 
in warm clouds (triangles) and 
cold clouds (circle) cloud are shown, based on observations of $\zeta$ Oph. 
(The apparent overabundance of S in cold gas is probably the result of
measurement uncertainties.)
The vertical bars represent the range of abundances found for the 
LIC (Table 1);
the asterisks are the $\epsilon$ CMa LIC values.
Since over twelve interstellar 
absorption components are seen towards $\zeta$ Oph, these abundances are guides only to the 
differences found in abundance patterns.  Not shown is the ratio Na/H=--6.64
in the cold cloud towards $\zeta$ Oph, where the ionization potential
of Na is 5.1 eV.  Elements with FIPs less than
13.6 eV (the hydrogen ionization potential) will be ionized in neutral
interstellar clouds.  Both abundance patterns and ionization must
be considered when estimating the abundances of neutrals in interstellar
gas.  These abundances, and the
reference abundances, are taken from Savage and Sembach (1996),
York (1983), and Table 1.
}
\label{abund.eps}
\end{figure}

Table 1 also summarizes Fe and Mg abundances towards several nearby stars.
A puzzling pattern emerges from the data in Table 1.  The stars
$\alpha$ CMa, $\epsilon$ CMa and $\alpha$ Cen, appear to have enhanced
Mg$^{+}$/H$^{\circ}$ and Fe$^{+}$/H$^{\circ}$ abundances in the foreground interstellar gas when
compared to the other stars in the table, a property which may be
partly attributed to ionization effects which are not included in the 
depletion estimates.  Of the stars listed in Table 1,
these three stars sample clouds in the most compact region of space around the 
solar system.  For stars sampling more distant regions, uncertainties 
could be introduced 
by to unresolved velocity structure in saturated lines, but the good
correlation between Fe$^{+}$ and Mg$^{+}$ suggests this is not the case
(\cite{frissi}).
Alternatively, this variation could be evidence for abundance variations in the gas-phase ISM over the longer sightlines within 20 pc of the Sun
(e. g.  Lallement et al. 1995, \cite{pis97,valwel,fr95}).  This variation
corresponds to a factor of $\sim$4 variation in gas phase abundances,
but much smaller variations in dust mass due to the highly
depleted nature of refractories.

\section{Hydrogen ``Wall'' at Heliosphere Nose\label{wall}}

The relative motions of the Sun and surrounding
interstellar cloud result in a pile-up of interstellar H$^{\circ}$ at the
nose of the heliosphere caused by the charge exchange coupling of 
of interstellar H$^{\circ}$ and H$^{+}$ (e. g. \cite{baranov1,baranov2,gay}, and references therein). This pile-up leaves a
signature in the Ly$\alpha$ absorption line which can, in principle,
be distinguished and modeled in the spectrum of nearby stars
(\cite{linwood97,frbaas}).  Gayley et al. (1997) have modeled the
Ly$\alpha$ due to the pile-up in the spectrum of $\alpha$ Cen,
and found that a barely subsonic heliosphere model with a Mach number of 0.9 provided
the best fit.  The model parameters used in this fit were n(H$^{\circ}$)=0.14 cm$^{-3}$,
n(e$^{-}$)=0.1 cm$^{-3}$, T=7,600 K, and V=--26 km s$^{-1}$.
For this set of parameters, if the excess velocity (over the sound
speed) is the Alfven speed resulting from an interstellar magnetic field,
then a magnetic field strength of B=3.1 $\mu$G is required to achieve
a Mach number of 0.9.  This gives
sound and Alfven speeds of 10.5 and 20.9 km s$^{-1}$ respectively.
Although the full range of parameter space has not yet been explored
for the modeling process, supersonic and subsonic models corresponding
to Mach numbers 1.5 and 0.7, or magnetic field strengths of 2 and 5
$\mu$G, were found to provide unacceptable fits to the observed Ly$\alpha$ profile.
Based on this exercise, evidently the properties of astrospheres may be
used to determine the pressure of the interstellar medium at the location
of a star, as predicted earlier by Frisch (1993).

\section{Concluding Remarks}

One new conclusion presented here is that in principle the
{\it in situ} pickup ion data can help resolve the 
outstanding question of whether the correct reference abundances for the LIC are
given by solar versus B-star abundances. 

A second new result is that the uncertainties on the LIC velocity
vector indicate that it is not yet clear whether the Sun and $\alpha$
Cen are immersed in the same interstellar cloud.

Based on the discussions in this paper, the best values for LIC properties are given by 
n(H$^{\circ}$)=0.22$\pm$0.06 cm$^{-3}$, n(e$^{-}$)=n(H$^{+}$)=0.1 cm$^{-3}$,
T=6,900 K and a relative Sun-cloud velocity of 25.8$\pm$0.8 km s$^{-1}$.
However, radiative transfer considerations in the LIC suggest that
the quoted neutral density is a lower limit.
Ulysses and EUVE observations of He$^{\circ}$ indicate a cloud temperature of T=6,400 K.
The magnetic field strength is weakly constrained to be in the
range of 2--3 $\mu$G.  Models of the Ly$\alpha$ absorption line towards $\alpha$
Cen are consistent with an Alfven velocity of 20.9 km s$^{-1}$,
which in turn is consistent with an interstellar magnetic field of 3 $\mu$G
in the absence of additional unknown contributions to the interstellar 
pressure.  
Ulysses and EUVE observations of interstellar He$^{\circ}$ within the solar
system give an upwind direction for the ``wind'' of
interstellar gas through the solar system, in the rest frame of the Sun,
of V=--25.9$\pm$0.6 km s$^{-1}$ arriving from the galactic direction
l=4.0$^{\circ}$$\pm$0.2$^{\circ}$, b=15.4$^{\circ}$$\pm$0.6$^{\circ}$.
Removing solar motion from this vector gives an
upwind direction for the LIC cloud in the LSR of V=--18.7$\pm0.6$ km s$^{-1}$
arriving from the direction l=327.3$^{\circ}$$\pm$1.4$^{\circ}$, b=0.3$^{\circ}$$\pm$1.0$^{\circ}$.

Through a combination of observations and theory,
uncertainties in the LIC electron density are narrowing.
Radiative transfer in the sightlines towards nearby stars require that
cloud models must be combined with data in order to deduce properties at the
cloud location.
Radiative transfer models of ionization in the LISM show interesting results, but
additional understanding of the input radiation fields is needed.
The Local Fluff complex is structured and inhomogeneous. 
Striking progress would be made in understanding this structure if 
interstellar absorption lines could be observed at resolutions of $\sim$1 km s$^{-1}$ in the
ultraviolet.  
The most glaring uncertainty is the absence
of detailed knowledge about the interstellar magnetic field.
Many of the most abundant elements in the LIC are ionized, and
densities of neutral atoms with FIPs less than 13.6 eV are
typically down by 1--3 orders of magnitude from the dominant ions.
The current approach of trying to understand the interaction of the ISM with the heliopause, from both the outside in and the inside out, is
finally bearing fruit.  

\section{Acknowledgements}

I would like to thank John Vallerga and a second ``anonymous'' referee
for thoughtful comments which have improved the quality of this paper.
I would also like to thank Dan Welty and Adolf Witt for helpful discussions.
I gratefully acknowledge the support of NASA grants NAG5-6188 and NAG5-6405.

%
%

\end{document}